# Channel characterization in screen-to-camera based optical camera communication

Vaigai Nayaki Yokar , Optical Communications Research Group, Department of Mathematics, Physics and Electrical Engineering, Northumbria University, Newcastle Upon Tyne, United Kingdom,
Hoa Le-Minh , Optical Communications Research Group, Department of Mathematics, Physics and Electrical Engineering, Northumbria University, Newcastle Upon Tyne, United Kingdom,
Zabih Ghassemlooy , Optical Communications Research Group, Department of Mathematics, Physics and Electrical Engineering, Northumbria University, Newcastle Upon Tyne, United Kingdom, *and*
Wai Lok Woo, Department of Computer and Information Sciences, Northumbria University, Newcastle Upon Tyne, United Kingdom.

*Abstract*—With the increase in optical camera communication (OCC), a screen to camera-based communication can be established. This opens a new field of visible light communication (VLC) known as smartphone to smartphone based visible light communication (S2SVLC) system. In this paper, we experimentally demonstrate a S2SVLC system based on VLC technology using a smartphone screen and a smartphone camera over a link span of 20 cms. We analyze the Lambertian order of the smartphone screen and carry out a channel characterization of a screen to camera link-based VLC system under specific test conditions.

## I. INTRODUCTION

Visible light communication (VLC) is a wireless communication technology that uses visible light as a medium for transmitting data [1] [2] . It utilizes light-emitting diodes (LEDs) to transmit data, which are modulated to carry information. VLC has gained significant attention in recent years due to its various advantages over traditional wireless communication technologies such as Wi-Fi, Bluetooth, and cellular networks [3] . VLC over screen to camera link has become a promising technology for short link communication on smartphone. The ever-increasing demand for high data-rate wireless communications has initiated a rapid progress in the research and development in the VLC based technology [4] . With the increased attention in the optical camera communication (OCC) can leverage the existing camera with light emitting diodes to establish a device-to-device communications, which is best known as smartphone-to-smartphone based visible light communication (S2SVLC) system or screen to camera link based optical communication system .

The research work in VLC for S2SVLC systems have been reported in the literature. This study explores the feasibility of using screen-to-camera communication for mobile payments [5]. The authors developed a prototype system and evaluated its performance in terms of accuracy, speed, and user experience. The study found that screen-to-camera communication is a viable technology for mobile payments and has several advantages over other technologies such as NFC. This study proposes a visual MIMO (multiple-input multiple-output) system for screen-to-camera communication in mobile devices. The system uses multiple LEDs on the screen and multiple photodiodes on the camera to transmit and receive data [6]. The study evaluates the performance of the system in terms of data rate, bit error rate, and user experience. The results show that the visual MIMO system is a promising technology for screen-to-camera communication in mobile devices. This study proposes a screen-to-camera communication system using colored lights. The system uses a set of colored lights on the screen to transmit data and a color filter on the camera to receive the data [7] [8] [9]. The study evaluates the performance of the system in terms of data rate, distance, and user experience. The results show that the system is a feasible technology for screen-to-camera communication in mobile devices. This study proposes a screen-to-camera communication system using barcode-like signals [7]. The system uses a set of rectangular bars on the screen to transmit data and an image processing algorithm on the camera to decode the signals. The study evaluates the performance of the system in terms of data rate, distance, and user experience. The results show that the system is a promising technology for screen-to-camera communication in mobile devices.

## II. SYSTEM DESIGN

The schematic diagram of a S2SVLC system is depicted in Fig. 1. At the Tx, a stream of data (text or any media format) is first converted into a binary stream (0's and 1's) and then encoded into an image for display on the smartphone screen. We noticed that (i) each bit intensity modulates a single or a group of pixels on the mobile smartphone screen; and (ii) pixels representing 0 or 1 bits are assigned with the white and black colors, respectively . At the Rx, a smartphone rear view camera is used to capture the transmitted image. The front camera can also be used to capture but with reduced capacity. Following machine learning-based algorithm the images are converted into a binary stream, where the white and black colored pixels are assigned the value of 0's and 1's accordingly [10]. The regenerated binary data is then converted into a text, see Fig. 2.

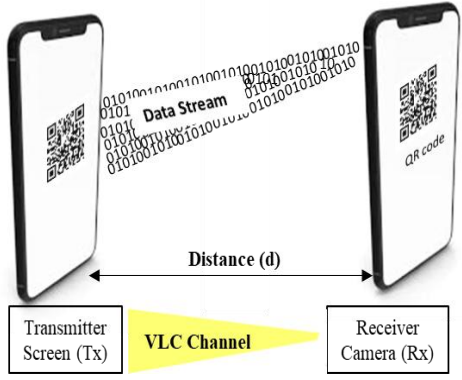

**Fig. 1.** Smartphone based VLC system (Tx and Rx pair).

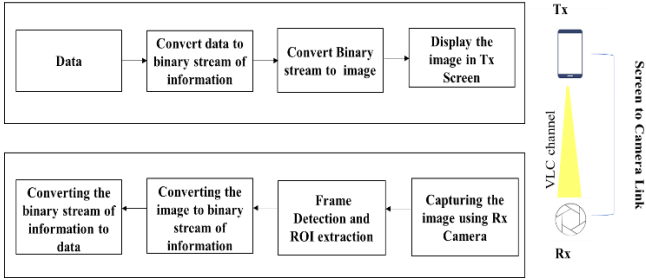

**Fig. 2.** System block diagram of S2SVLC.

Currently, detection of QR and ASCII codes is carried out by means of an internal decoder that uses classical computer vision algorithms. For most cases, the above scheme works well, but can have delays during the transmission of large data frames consisting of up to 4000 characters of data. In S2SVLC [9], the delay becomes increasingly noticeable during multi-frame transmission of a large file [11] [12]. To avoid these problems and improve speed, a supervised learning model is developed to detect and classify the QR and ASCII codes.

Fig. 2. shows the schematic block diagram of the proposed VLC system. A typical on-off keying (OOK)-based S2SVLC system consist of two devices (Google pixel 6 pro) that are utilized as the Tx and the Rx for displaying and collecting the data frames respectively. Each frame is made up off $M \times N$ pixels [10]. In OOK or system that use colour shift keying (CSK), to assign black for '1' and white of '0' pixel value. The OOK is the familiar modulation technique in S2SVLC system. At the Tx, the data stream $d_s(t) \in \{1,0\}$ is converted to a $200 \times 200$ sized data frames $d_f(t)$ (i.e., an ASCII character in byte form), where each frame is made up of an image that has been divided into $M \times N$ pixels, packed with 182 bits respectively. The intensity modulated light $P(t)$ is transmitted in an indoor environment through the free space and is captured on the Rx side using the camera. The channel DC gain for the receiver located at a distance of $d$ and angle ø with respect to the with respect to the Tx can be approximated as:

$$H_{los}(0) = \begin{cases} \frac{A_r(m_1+1)}{2\pi d^2} cos^m(\emptyset)T_s(\psi)cos\psi & 0 \leq \psi \leq \psi_c \\ 0 & elsewhere \end{cases} \quad (1)$$

Where $A_r$ is the image displayed in the Tx screen and it depends on the distance $d$ between the Tx and the Rx, g(ψ) and $T_s(\psi)$ are the gains of the optical conductor and optical filter respectively. The radiance angle is denoted by ø, ψ represents the incidence angle, $\psi_c$ represents the field of view (FOV) semi angle of the camera and m represents the Lambertian order of emission of the Tx. For the line of sight (LOS) link, the received signal is given by:

$$P_{r\text{-}los} = H_{los}(0)P_t . \quad (2)$$

The order of Lambertian emission $m$ of the Tx is related to the LED semi angle at half power $\emptyset_{1/2}$ by

$$m = \frac{-\ln 2}{\ln(cos\emptyset_{1/2})}. \quad (3)$$

The Lambertian radiant intensity is expressed as:

$$R(\emptyset) = \frac{(m+1)}{2\pi} cos^m(\emptyset) \quad (4)$$

The detailed model of the proposed link is represented in Fig. 3. Where, the $d$ between the Tx and Rx is 20 cm and ø is varied over a range of $0 - 180°$. Gaussian noise is evenly distributed over the signal. This means that each pixel in the noisy image is the sum of the true pixel value and a random distributed noise value. The gaussian distribution can be approximated as:

$$p(x) = \frac{1}{\sigma\sqrt{2\pi}} e^{-\frac{1}{2}\left(\frac{x-\mu}{\sigma}\right)^2}, \quad (5)$$

Where σ is the standard deviation and μ is the mean.

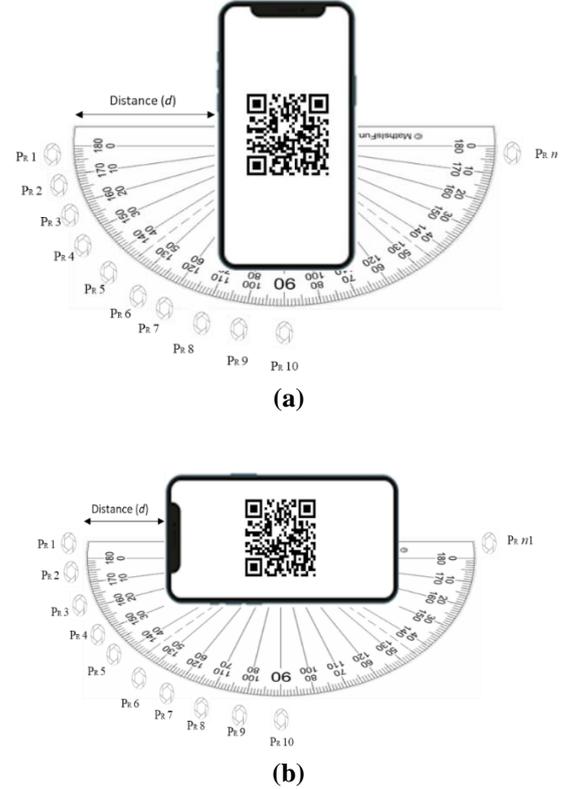

**Fig. 3.** Configuration of the S2SVLC system under: (a) portrait, and (b) landscape conditions.

Here, we have assumed that the camera is directly facing the Tx with no tilting or rotation angle. Note, (*i*) distance (*d*) between the Tx and the Rx to constant of 20 cm; (*ii*) The experimental investigation is carried out in an no light environmental condition. and (*iii*) the power(uW) is measured across the field of view (FOV) of the Tx screen in both landscape and portrait condition as represented in Fig.3.

III. RESULTS AND DISCUSSION

We have used the smartphone (Google pixel 6 pro) placed on the same axis to communicate with each other as depicted in Fig.4. The image is captured at a link span of 20 cm, no tilt, no rotation and over the FOV range measuring up to 0-180° at no illumination condition. The key parameters adopted in the work are represented in Table I.

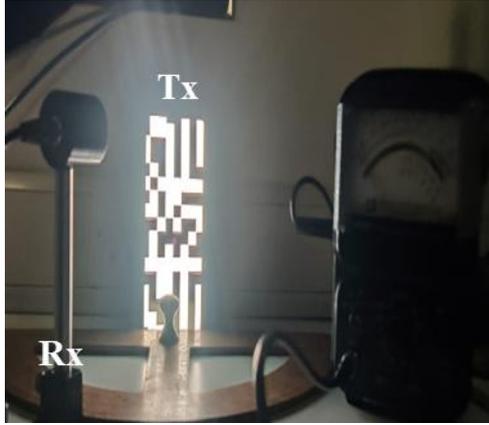

(a)

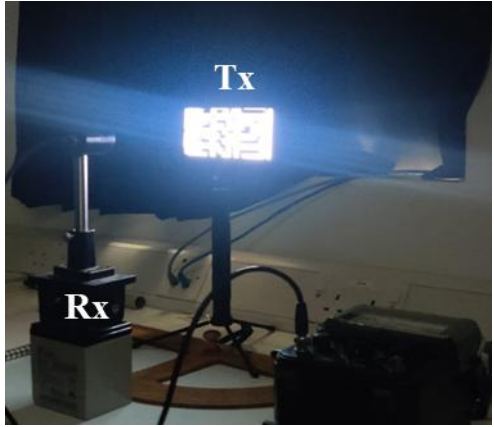

(b)

**Fig. 4.** The experimental setup for the S2SVLC system under: (a) portrait, and (b) landscape conditions.

TABLE I : KEY SYSTEM PARAMETERS

| Parameter | Value |
|---|---|
| No. of camera pixels | 50 MP |
| Aperture | f/1.85 |
| Sensor type | CMOS, Laser detection autofocus (LDAF) |
| Sensor model | Sony IMX386 |
| Sensor size | 1/1.31" wide |
| Camera frame rate | 60 fps |
| Camera focus | Auto focus |
| Tx display | OLED |
| Tx display size | 6.41 inch |
| Tx frame rate | 60 fps |
| Image size (cell $x$ cell) | 200 x 200 |
| Distance between the Tx and the Rx ($d$) for experimental investigation | 20 cm |
| Tilt angles of the Rx with respect to the Tx for experimental investigation | 0° |
| Rotation angles of the Rx with respect to the Tx for experimental investigation | 0° |

The power received is measured over the range of 0 - 180°. Note, that each pixel in the noisy image is the sum of the true pixel value and a random distributed noise value. The power is evenly distributed over the signal in a gaussian distribution and represented in Fig. 5.

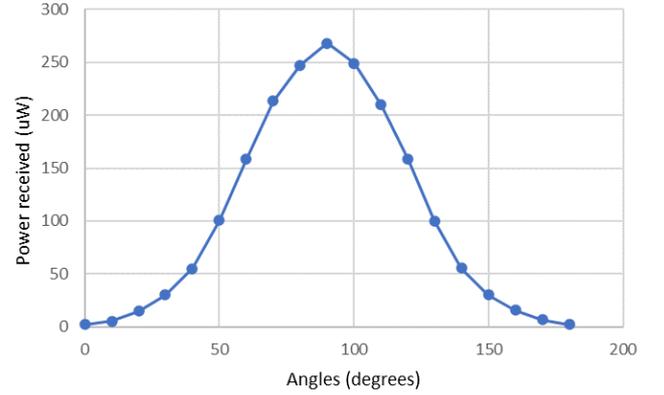

(a)

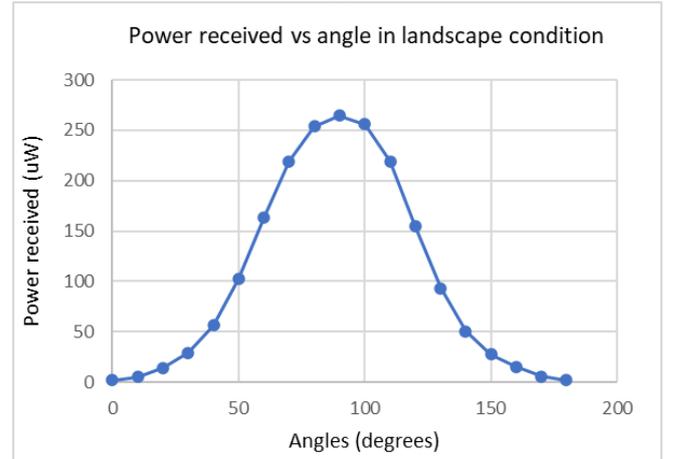

(b)

**Fig. 5.** The power distribution for the S2SVLC system under: (a) portrait, and (b) landscape conditions.

For the gaussian distribution over the FOV, we measure the beam profile of the Tx, which is close to Lambertian as in (3). We normalized the beam profile and by fitting a Lambertian curve to the beam profile we obtained Lambertian order $m$ of 1 as represented in Fig. 6.

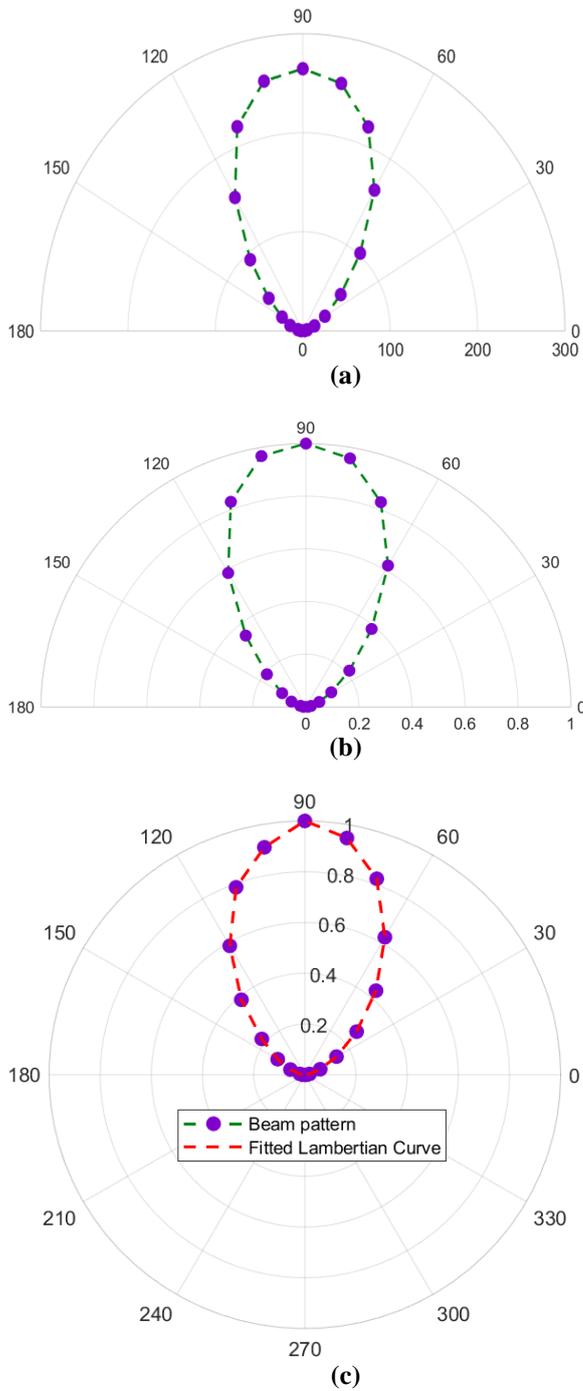

**Fig. 6.** The measured beam profile for the S2SVLC channel (a) beam profile (b) normalized beam (c) fitted Lambertian beam profile.

A short message of 'optical communications research group' was transmitted over the link. At the Rx, the image was decoded back to text using the camera with a characteristic as represented in Table I. Note, as was observed the longer the link spans the smaller will be the size of the captured image. The ROI is extracted in the Rx end and is rescaled to match the Tx image size. Consequently, beyond the transmission range the with the light intensity lowering from the transmission end the link between the Tx and Rx breaks. Fig.7. shows the success rate of the data transmission against the link span up to 55 cms. At a link span of 40cm, the success rate reduces to 98%, the success rate of the received bits reduces as the link span increases. This is due to the fact that, the received power is spread over a number of pixels, hence lower SNR.

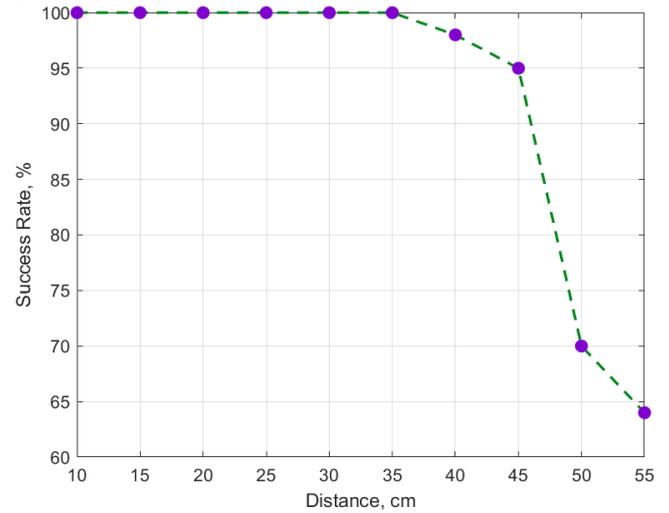

**Fig. 7.** Success Rate of the data transmission over a link span for the S2SVLC transmission.

This delves into the key characteristics and performance metrics of smartphone screen used as Tx in S2SVLC communication system. It examines various aspects such as screen spectral analysis, beam profiling, and Lambertian order estimation to understand their impact on the S2SVLC performance.

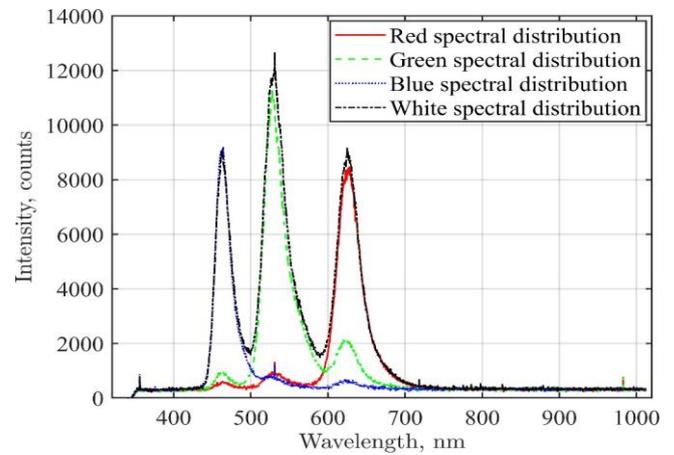

**Fig. 8**. Smartphone spectral distribution.

Beam profiling is a critical aspect in the field of optical technology, it involves measuring the intensity of the light beam at different points across its cross-section. This provides a detailed picture of beam's shape, size, and uniformity. In S2SVLC communication, the screen beam profiling provides crucial information which helps with improved data transmission, enhancing screen reliability, optimizing receiver design and assessing system performance. The screen beam profiling [13] is carried out for a white screen with all pixels activated, which is made by combinations of red, green and blue spectrum. There are two

different beam profiling techniques carried out for the analysis [14]:
1. Scanning Slit Profilers: Profiler devices utilize a narrow slit to scan across a beam. In the case of a scanning slit profiler, the device functions by directing a beam through a narrow slit or a series of slits. As the slit traverses across the beam, it samples a small portion of the beam's cross-section sequentially. Subsequently, the light passing through the slit is detected by a photodetector, which measures the intensity of the beam at each position [15].
2. Knife-Edge Technique: The knife-edge technique entails the movement of a sharp edge, known as the "knife," through the beam while measuring the reduction in power as the knife-edge obstructs more of the beam. Through analysis of the rate at which the power diminishes as the edge traverses through the beam, one can deduce the spatial profile of the beam [16].

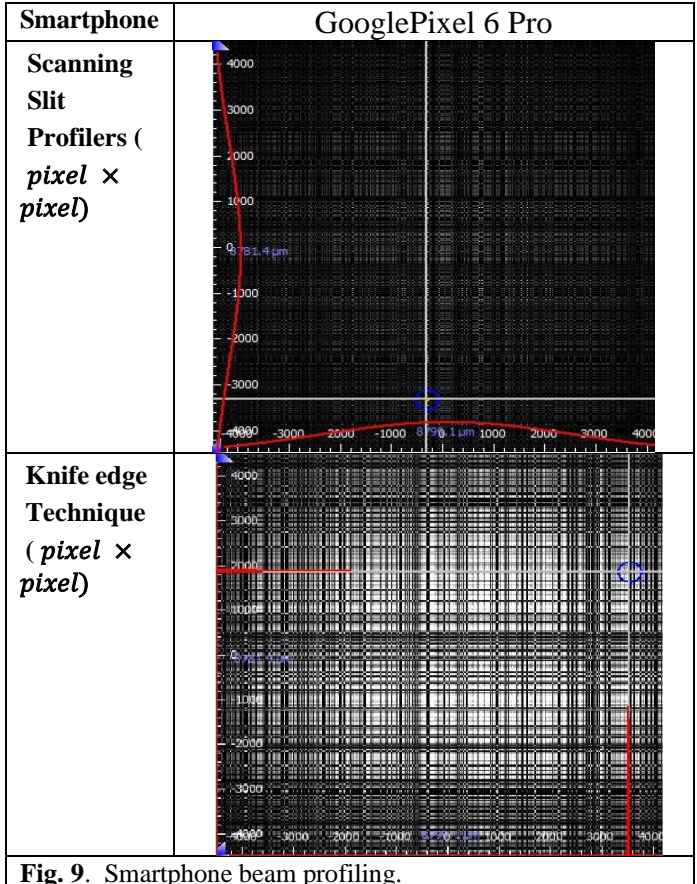

| Smartphone | GooglePixel 6 Pro |
|---|---|
| Scanning Slit Profilers ($pixel \times pixel$) | |
| Knife edge Technique ($pixel \times pixel$) | |

**Fig. 9**. Smartphone beam profiling.

The scanning slit profilers display a high resolution of the beam's intensity distribution along two orthogonal axes. The red line likely represents a cross-section of the beam intensity, showing the distribution along one axis. The knife-edge delineation of specific data points or regions of interest, depicted by the red and blue lines, is indicative of measurements pertinent to the beam's width or other parameters crucial for optimizing the light source [14].

The beam profile exhibits a high degree of uniformity in the central region, as evidenced by the flatness of the top portion of the curve. This indicates that the light intensity is relatively consistent across the core of the beam, which is crucial for minimizing signal distortion and ensuring efficient data transmission in S2SVLC communication. The smooth transition of the curve suggests a Gaussian-like beam profile, indicative of good beam quality. The peak intensity, determined by the maximum value on the y-axis, is indicative of the signal strength, directly influencing the communication range and data rate. The sharp transitions at the edges of the beam define the core region and provide an estimate of the beam width, which is a critical parameter for optimizing the alignment and focusing of the receiver.

## IV. CONCLUSION AND FUTURE WORK

We have reported the investigation of a S2SVLC system using real smartphone under indoor conditions. We determined the Lambertian order of the smartphone by carrying out experimental investigation of the smartphone in both landscape and portrait position. The success rate of the data transmission over a link span of 10-55 cm was also carried out.

REFERENCES

[1] V. N. Yokar, Hoa-Le-Minh, F. Ghassemlooy, and W. L. Woo, "A Novel Blur Reduction Technique For QR And ASCII Coding In Smartphone Visible Light Communications," in *2022 13th International Symposium on Communication Systems, Networks and Digital Signal Processing (CSNDSP)*, Jul. 2022, pp. 428–433. doi: 10.1109/CSNDSP54353.2022.9907993.
[2] Y. Almadani *et al.*, "Visible Light Communications for Industrial Applications—Challenges and Potentials," *Electronics*, vol. 9, no. 12, 2020, doi: 10.3390/electronics9122157.
[3] Z. Ghassemlooy, W. Popoola, and S. Rajbhandari, *Optical Wireless Communications: System and Channel Modelling with MATLAB*. CRC Press, 2019. [Online]. Available: https://books.google.co.uk/books?id=uSlNvgAACAAJ
[4] S. Shen *et al.*, "Unified monitoring and telemetry platform supporting network intelligence in optical networks," *J. Opt. Commun. Netw.*, vol. 17, no. 2, pp. 139–151, Feb. 2025, doi: 10.1364/JOCN.538552.
[5] T. Hao, R. Zhou, and G. Xing, "COBRA: Color barcode streaming for smartphone systems," presented at the Proceedings of the 10th international conference on Mobile systems, applications, and services, 2012, pp. 85–98.
[6] Q. Wang, M. Zhou, K. Ren, T. Lei, J. Li, and Z. Wang, "Rain Bar: Robust Application-Driven Visual Communication Using Color Barcodes," in *2015 IEEE 35th International Conference on Distributed Computing Systems*, Jul. 2015, pp. 537–546. doi: 10.1109/ICDCS.2015.61.
[7] M. Stafford, A. Rogers, S. Wu, C. Carver, N. S. Artan, and Z. Dong, "TETRIS: Smartphone-to-Smartphone Screen-Based Visible Light Communication," in *2017 IEEE 14th International Conference on Mobile Ad Hoc and Sensor Systems (MASS)*, Oct. 2017, pp. 570–574. doi: 10.1109/MASS.2017.101.
[8] V. N. Yokar, H. Le-Minh, Z. Ghassemlooy, and W. L. Woo, "Performance evaluation technique for screen-to-camera-based optical camera communications," *IET Optoelectronics*, vol. n/a, no. n/a, Aug. 2023, doi: 10.1049/ote2.12102.
[9] V. N. Yokar, H. Le-Minh, Z. Ghassemlooy, and W. L. Woo, "Data Detection Technique for Screen-to-Camera Based Optical Camera Communications," presented at the 2024 14th International Symposium on Communication Systems, Networks and Digital Signal Processing (CSNDSP), IEEE, 2024, pp. 233–237.
[10] V. N. Yokar *et al.*, "Smartphone Beam Profile in a Screen-to-Camera-Based Optical Communication System," in *2023 17th International Conference on Telecommunications (ConTEL)*, Jul. 2023, pp. 1–6. doi: 10.1109/ConTEL58387.2023.10199032.
[11] V. N. Yokar, H. Le-Minh, L. N. Alves, S. Zvanovec, W. L. Woo, and Z. Ghassemlooy, "Non-Blind Image Restoration Technique in Screen–to–Camera based Optical Camera Communications," presented at the 2024 7th International Balkan Conference on Communications and Networking (BalkanCom), IEEE, 2024, pp. 101–106.

[12] V. Yokar *et al.*, "Fast Link Recovery via PTP-synchronized Nanosecond Optical Switching," *arXiv preprint arXiv:2412.13778*, 2024.
[13] L. Hatanaka and L. Gragnic, "Beam profilers," *Photoniques*, no. 119, pp. 68–72, 2023.
[14] J. Hue, J. Dijon, P. Garrec, G. Ravel, L. Poupinet, and P. Lyan, "Beam characterization: application to the laser damage threshold," presented at the Laser-Induced Damage in Optical Materials: 1998, SPIE, 1999, pp. 633–644.
[15] C. B. Roundy, "Current technology of beam profile measurements," *OPTICAL ENGINEERING-NEW YORK-MARCEL DEKKER INCORPORATED-*, vol. 70, pp. 349–422, 2000.
[16] W. Plass, R. Maestle, K. Wittig, A. Voss, and A. Giesen, "High-resolution knife-edge laser beam profiling," *Optics communications*, vol. 134, no. 1–6, pp. 21–24, 1997.